\documentclass[aps,prl,twocolumn,showpacs,floatfix]{revtex4}
\usepackage{epsfig}
\begin{document}

\title{Thin Ohmic or superconducting strip with an applied ac
       electric current}

\author{E.~H.~Brandt}
\affiliation{Max-Planck-Institut f\"ur Metallforschung,
   D-70506 Stuttgart, Germany}

\date{\today}

\begin{abstract}
The complex impedance,  currents, and electric and
magnetic fields are calculated as functions of resistivity
and frequency or London depth for a long thin strip with
applied ac current. Both Ohmic and superconducting strips
are considered. While the inductance per unit length of
the strip {\em depends} on the strip length logarithmically,
the sheet current, magnetic field, resistance, and magnetic
susceptibility are {\em independent} of this length.
It is found that the enhancement of resistance by the
skin effect in thin Ohmic strips is much weaker
(logarithmic) than in thick wires.
\end{abstract}


\maketitle


The distributions of the electric and magnetic fields inside
and around a long wire carrying an applied ac current $I_a$ is
a standard problem in electrodynamics when the wire has circular
cross section \cite{1}. In this case, the solutions of Maxwell
equations away from the current contacts depend only on the
radial coordinate $r$ and are easily obtained for both
{\em Ohmic} wires with resistivity $\rho$ and  superconducting
wires in the {\em Meissner state} with London penetration
depth $\lambda$. However,
when the wire has a non-circular cross section, e.g., for
flat strips, the problem becomes difficult. While the theory
of thin strips in an applied ac field with no applied current
is known since long time
-- it is simpler since the strip length drops out from all
final results -- apparently, no analytic or transparent
numerical solutions are available for the even more important
response of thin strips to applied ac current. This response
is required, e.g., to understand
SQUID systems \cite{2} or to analyze the important experiments
on temperature dependent magnetic field profiles of
superconducting strips \cite{3,4}.
 In particular, the dependences of inductance,
resistance, and field and current distributions on $\lambda$,
$\rho$, ac frequency, {\em and strip length\/}, are of principal
interest but are still unknown.

  In this paper we settle this fundamental question and obtain
these dependences for a current-carrying thin strip
of width $2a$, length $l \gg a$, and thickness $d \ll a$,
filling the volume $|x| \le a$, $|y| \le d/2$, $|z| \le l$.
In this basic case, the magnetic field around the strip
depends only on the sheet current
$J(x) = \int_{-d/2}^{d/2} j(x,y)\, dy$, which like the
current density $j$ and total current $I_a$ flows along $z$.
Interestingly, the inductance per unit length of a long
strip depends on its length $l$ although the field
and current distributions in the strip and its resistance
and magnetic susceptibility are independent of $l$.

  This strip example has the additional advantage that the
nontrivial magnetic field distribution at the surface of
the conductor can be measured by magneto-optics using flat
indicator films or linear arrays of Hall probes \cite{3,4}.
Note that with cylindrical wires there is no analog to such
field profiles since the wire surface $r=a$ exhibits
constant and material-independent field magnitude $I_a/2\pi a$.

  To solve the strip problem with minimum effort we
derive an equation of motion for the sheet current $J(x,t)$,
which in general is driven by a time-dependent perpendicularly
applied magnetic field $H_a(x,t) \| y$  and/or an electric
current $I_a(t)$ applied at the ends of the strip, far away
from the considered section. This method implicitly accounts for
the complicated magnetic field around the strip, which thus
does not have to be cut off or approximated as in other
numerical methods. For strips with no applied current in
a perpendicular $H_a(t)$,
integral equations for $J(x,t)$ and $j(x,y,t)$ were obtained
in Refs.~\onlinecite{5} for thin and thick strips with linear
or nonlinear resistivity, assuming zero London depth
$\lambda=0$. But actually, from the linear results of
Refs.~\onlinecite{5} one can obtain the equations and results
also for a superconductor strip in the Meissner state with
arbitrary $\lambda$, namely, by replacing the Ohmic (real)
resistivity $\rho$ by an imaginary value
$\rho(\omega) = i\omega\mu_0\lambda^2$. The circular frequency
$\omega$ then drops out from the final (static) results.

  General equations for thin and thick strips with both applied
$H_a$ and $I_a$ and with finite $\lambda$ were derived in
Ref.~\onlinecite{6} using the rather general voltage-current
law valid inside superconductors with or without vortices,
  \begin{eqnarray}  
  E = E_v({\bf r},j,{\bf B}) + \mu_0 \lambda^2({\bf r})
      \partial j / \partial t \,.
  \end{eqnarray}
In it $E_v({\bf r},j,{\bf B}) = \rho_v({\bf r},j,{\bf B}) j$ is
the electric field caused, e.g.,  by moving vortices, and
${\bf B}$ is the magnetic induction. The second
(London) term describes the acceleration of the massive charge
carriers (Cooper pairs) by the electric field.
The flux-flow term $E_v$ can be linear or nonlinear,
e.g., $\rho_v =\rho_{FF} ={\rm const}\cdot B$ for free flux flow,
or $\rho_v \propto |j|^{n-1}$ for thermally activated depinning
of vortices, where $n \gg 1$ is the flux-creep exponent.

   For the particular case of a thin strip with applied current
$I_a$ but no applied field $H_a$ (or with large dc background
field), the sheet current has the symmetry
$J(-x) = J(x)$. The equation for $J(x,t)$ then reads \cite{6}
  \begin{eqnarray}  
  {\partial J(x,t)\over \partial t} = \mu_0^{-1}\!\! \int_0^a \!\!
  dx' K(x,x') \,[E_a(t) - E_v(x',t)]\,.
  \end{eqnarray}
Here $E_v(x',t) =E_v(x',j,B)$ with $j= J(x',t)/d(x')$ and
$B = \mu_0 H(x',t)$ the magnetic induction in the strip.
The spatially constant electric field $E_a(t)$ (along $z$, like
$j$ and $J$) formally drives the currents; it will drop out
later when the solution is expressed in terms of $I_a(t)$.
Finally, the integral kernel $K(x,x')$ is the inverse of a kernel
$K^{-1}(x,x')$ and is defined by
  \begin{eqnarray}   
  \int_0^a \!\! dx'' K(x,x'')\, K^{-1}(x'',x')=\delta(x-x'),
  \nonumber \\[-2mm]
  K^{-1}(x,x') = {1\over 2\pi} \ln{l^2\over |x^2-x'^2|}
    + \Lambda(x) \,\delta(x-x') \,,
  \end{eqnarray}
with $\Lambda(x) = \lambda^2(x) /d(x)$ the effective magnetic
penetration depth of a superconductor film with thickness
$d < \lambda$. Note that Eqs.~(2) and (3) apply also to
inhomogeneous strips with $\lambda$, $\rho_v$, and $d$ depending
on $|x|$ sufficiently smoothly (over length scale $\gg d$).
This property can be used to simulate edge currents \cite{3,4}
caused by a geometrical barrier \cite{7,8} for flux penetration.

For a strip with Ohmic resistivity $\rho$ and
applied ac current $I_a(t) = I_{a0} \exp(i\omega t)$ one has
$\Lambda = 0$ and $E_v = \rho_v j = (\rho_v/d) J$,
or one may formally put $E_v=0$ and
$\Lambda = -i\rho /(\mu_0\omega d)$.
Both methods yield the same equation for
$J(x,t)=J_0(x) \exp(i\omega t)$ or $E(x,t)=\rho J(x,t)/d$
driven by $E_a(t)=E_{a0} \exp(i\omega t)$,
  \begin{eqnarray}  
  J_0(x) = -{i\,E_{a0} \over\mu_0 \omega} \int_0^a \!
           dx'\, K(x,x') \,.
  \end{eqnarray}
The integral kernel $K(x,x')$, Eq.~(3), is now complex due to
the imaginary $\Lambda$. Dividing this by the
current amplitude  $I_{a0} =2 \int_0^a  dx\,J_0(x)$, the
electric field amplitude $E_{a0}$ drops out:
  \begin{eqnarray}  
  {J_0(x) \over I_{a0}} = \int_0^a \!\!\! dx' K(x,x') \bigg/
  2\! \int_0^a \!\!\! dx\!   \int_0^a \!\!\! dx' K(x,x')  \,.
  \end{eqnarray}
The magnetic field caused by this sheet current is
${\bf H}(x,y,t) ={\bf H}_0(x,y) \exp(i\omega t) =(H_x, H_y)$
with
  \begin{eqnarray}  
  {\bf H}_0(x,y) = \nabla\! \times {\bf\hat z}\!
  \int_{-a}^a \!\!\! dx'\,{J_0(x')\over 4\pi}\,
  \ln{l^2\over (x-x')^2+y^2}\,.
  \end{eqnarray}
   Interestingly, the profiles $J_0(x)$ and ${\bf H}_0(x,y)$
obtained numerically from Eqs.(4)-(6) {\it do not depend} on
the strip length $l$, though the kernel $K(x,x')$ looks
different for different ratios $l/a$.
This may be understood by physical arguments, but to prove
it mathematically appears to be difficult.
See also the similar theory for a double strip \cite{9}.
However, the  complex resistance or impedance $Z =i\omega L +R$
per unit length {\it does depend} on $l$ logarithmically
($R$ = real resistance, $L$ = real inductance).
Noting that $E_a l$ is the voltage drop along
the strip, one obtains
  \begin{eqnarray}  
  {Z\over l} = {E_{a0} \over I_{a0}} = i\omega \mu_0 \bigg/
  2 \! \int_0^a \!\!\! dx \int_0^a \!\!\! dx' K(x,x') \,.
  \end{eqnarray}
\begin{figure}  
\epsfxsize= .87\hsize  \vskip 1.0\baselineskip \centerline{
\epsffile{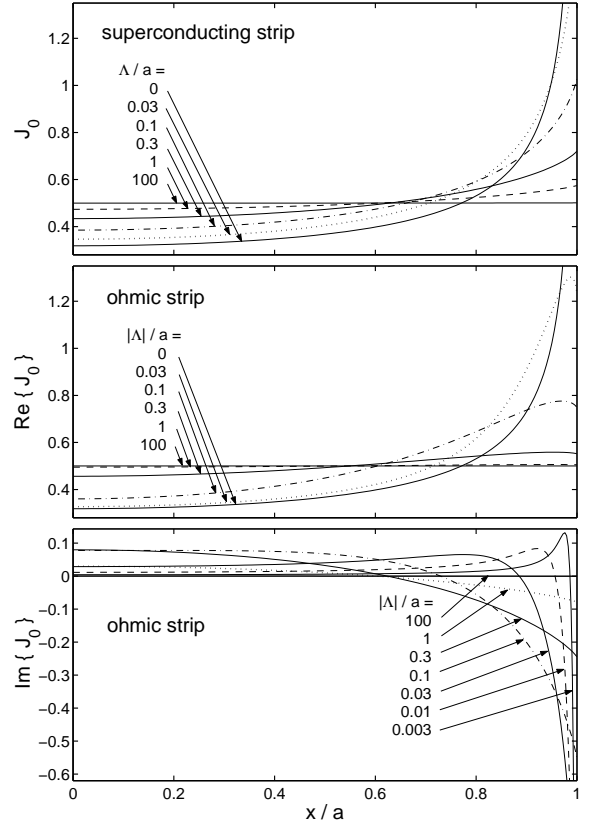}~~}  
\caption{\label{fig1}
   The complex amplitude of the sheet current $J_0(x)$ in
superconducting and Ohmic strips with applied ac current,
Eq.~(9).
Shown are the real (in-phase) and imaginary (out-of-phase)
components in units of $I_{a0} / a$ for various 2D penetration
depths $\Lambda$ or skin depth depths $\delta $
($|\Lambda| = \delta^2 / 2d$).}
\end{figure}        
\begin{figure}  
\epsfxsize= .87\hsize  \vskip 1.0\baselineskip \centerline{
\epsffile{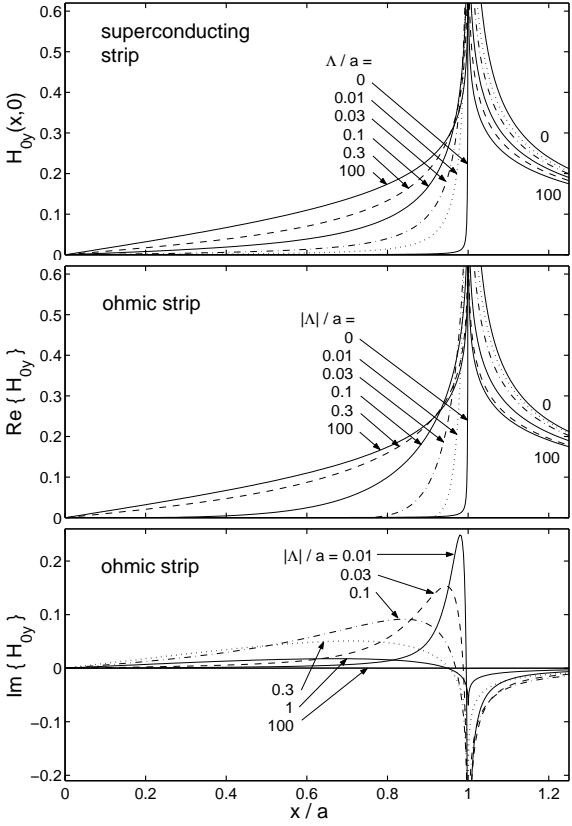}~~}  
\caption{\label{fig2}
   The complex amplitude of the perpendicular magnetic field
$H_{0y}(x,0)$ in the plane of superconducting and Ohmic strips
with applied ac current, Eq.~(10).
Shown are the real and imaginary parts in units of
$I_{a0} / a$ for various $|\Lambda|$.}
\end{figure}        

  We evaluate the integrals (2)--(7) using unit length $a=1$ and
introducing a grid $x_i$ ($0< x_i < 1$, $i=1,2,\dots,\,N)$
with weights $w_i$ such that for any sufficiently smooth
function $f(x)$ one has $\int_0^a\! f(x)\,dx \approx
\sum_{i=1}^N w_i f(x_i)$. A grid with grid points densely
spaced near the strip edge $x=1$, is obtained by introducing
an auxiliary grid $u_i = (i-{1\over 2})/N$ and then putting,
e.g., $x_i = (3u_i-u_i^3)/2$, $w_i = 3(1-u_i^2)/2$.
This amounts to a substitution of variables in the integral,
$\int_0^1\! f(x)\,dx = \int_0^1\! f[x(u)]\,x'(u)\,du$.
The integral kernel $K(x,x')$ now becomes a matrix \cite{5,6}:
  \begin{eqnarray}  
 K_{ij} &=& (w_j Q_{ij}+\Lambda(x_i)\,\delta_{ij})^{-1} \,,
   \nonumber\\
 Q_{ij} &=& {1\over 2\pi} \ln{l^2 \over|x_i^2 -x_j^2|}
   \,,~~i\ne j \,, \nonumber \\
 Q_{ii} &=& {1\over 2\pi} \ln{\pi\, l^2 \over x_i w_i} \,,
  \end{eqnarray}
where $\delta_{ij}$ equals $1$ when $i=j$ and 0 otherwise.
The diagonal term $Q_{ii}$ in Eq.~(8) is chosen such that
the numerical error
decreases with a high power of the grid number $N$, say, as
$N^{-2}$ or $N^{-3}$, depending on the chosen grid, while any
different choice gives a larger error $\propto N^{-1}$.
The current and field profiles are then obtained as
  \begin{eqnarray}  
  &J_0(x_i) =& {I_{a0} \over 2a} \sum_{j=1}^N K_{ij} \bigg/\!
  \sum_{k=1}^N \sum_{l=1}^N  w_k K_{kl} , \\
  &H_{0y}(x_i) =&  \nabla_i \sum_{j=1}^N  w_j Q_{ij}
  \,J_0(x_j)\,,
  \end{eqnarray}
where $\nabla_i$ means the numerical derivative $d/dx$;
this expression (10) is much more accurate than the direct
computation of $H_{0y}(x_i)$ from $J_0(x_i)$ by
Amp{\`e}re's Law using a matrix with singular terms
$\propto 1/(x_i-x_j)$.

The profiles $J_0(x)$ and $H_{0y}(x,0)$ for various $\Lambda$
are depicted in Figs.~1 and 2. For Ohmic and superconducting
strips one has the same limits \cite{10}:
at $|\Lambda| \ll a$ (ideal screening) one has
$J_0(x) =(I_{a0}/\pi)(a^2-x^2)^{-1/2}$,
$H_{0y}(x,0) = 0$ for $|x| < a$  and
$H_{0y}(x,0) =(I_{a0}/2\pi)\,{\rm sign}(x) (x^2 -a^2)^{-1/2}$
for $|x| >a$,
and at $|\Lambda| \gg a$ (weak screening)
$J_0(x) =(I_{a0}/2 a) =$ const  and
$H_{0y}(x,0) =(I_{a0}/2\pi)\, {\rm sign}(x)
\ln |( a -|x|) / (a +|x|) | $.
The imaginary (out-of-phase) components of $J_0(x)$ and
$H_{0y}(x,0)$ are zero in these limits, but are maximum
for Ohmic strips at some $x$ dependent value of
$|\Lambda| = \delta^2 /2 d$ where $\delta$ is the skin depth.

The complex impedance, Eq.~(7), is calculated as
  \begin{eqnarray}  
  Z = i\omega L +R = i \omega \mu_0 l  \bigg/\!
  \sum_{k=1}^N \sum_{l=1}^N  w_k K_{kl} \,.
  \end{eqnarray}
For superconducting strips this yields $R=0$ and the
inductance (partly obtained analytically)
  \begin{eqnarray}  
   L= {\mu_0 l \over 2\pi}\Big[ \ln{2l \over a} +
     \alpha\Big( {\Lambda\over a}\Big)
     \Big]  + {\mu_0 l \Lambda  \over 2a}
  \end{eqnarray}
with $\alpha(0)=0$ and $\alpha(\infty)= 3/2 -\ln4 = 0.11371$.
To good approximation, the first term is the geometric
inductance $L_m$ and the last term the kinetic inductance
$L_k$, while the small middle term
$\propto \alpha$ is shared by both $L_m$ and $L_k$,
which slightly depend on the current distribution, Fig.~1.
For Ohmic strips we obtain $Z =i\omega L +R$,
  \begin{eqnarray}  
   Z= i\omega {\mu_0 l\over 2\pi} \Big[ \ln {2l \over a} +
   \beta \Big( {|\Lambda| \over a}\Big) \Big] +
   R_{\rm strip} \gamma \Big( {|\Lambda| \over a} \Big)
  \end{eqnarray}
with $| \Lambda | = \rho/(\mu_0 \omega d) = \delta^2 /2d$
and $R_{\rm strip}=\rho l /2ad$. The functions
$\alpha$, $\beta$ and $\gamma$
are shown in Fig.~3. One has $\beta(0) = 0$ (ideal screening),
$\beta(\infty) = \alpha(\infty)= 3/2 -\ln4$ (uniform current),
$\gamma(\infty) = 1$, and for $|\Lambda| \ll a$,
$\gamma \approx 1+(2/\pi^2) \ln(0.14a/|\Lambda|)$ (skin effect).
The same $\gamma$ results from Ohmic dissipation,
$\gamma = 4a \int_0^a |J_0(x)/I_{a0}|^2 dx$, Eq.~(5).
Good fits for all $\Lambda$ are (see Fig.~3)
  \begin{eqnarray}  
  \alpha &\approx & \alpha(\infty)~ (\, \tanh[\,0.435 \ln
    (\, \Lambda/0.074\, a) \,] + 1) /2 \,, \\
  \beta &\approx & \alpha(\infty)\, \sqrt{(\tanh[\, 0.92
    \ln(|\Lambda| /0.15\, a) \,] + 1 ) /2 } \,, ~\\
  \gamma &\approx & 1 + (4/3\pi^2)\, \ln [\,1 +
  (0.14\, a / |\Lambda|)^{3/2} \,] \,.
    \end{eqnarray}

\begin{figure}  
\epsfxsize= .87\hsize  \vskip 1.0\baselineskip \centerline{
\epsffile{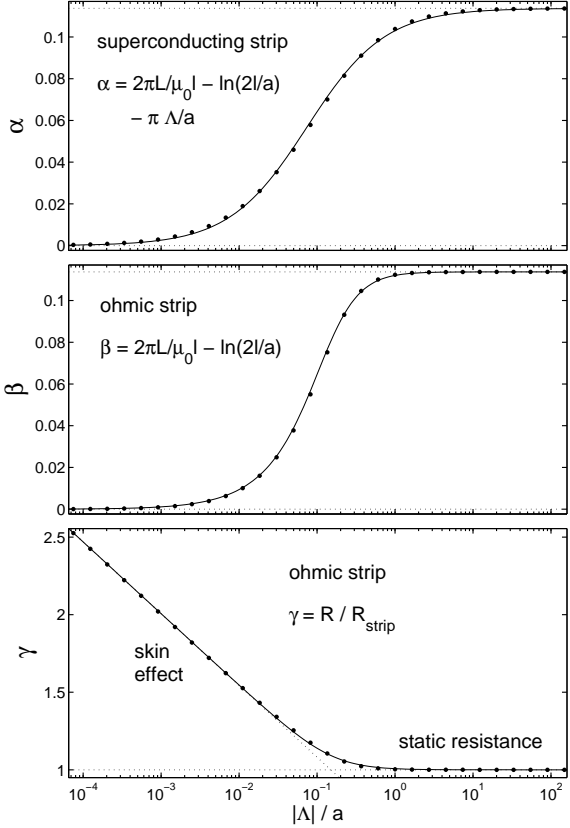}~~}  
\caption{\label{fig3}
  The functions $\alpha$, $\beta$, and $\gamma$ of $|\Lambda|/a$
entering the inductance and resistance of a strip,
Eqs.~(12) and (13) (dots). The solid lines show the
approximations Eqs.~(14)-(16).}
\end{figure}        

  For comparison we give here also the corresponding expressions
for a uniform round wire of radius $a$, with applied current
$I_a(t) = I_{a0} \exp(i\omega t)$, current density
$j(r,t) =j_0(r) \exp(i\omega t)$, and magnetic field
${\bf H} = \hat{\bf{\varphi}} H_0(r) \exp(i\omega t)$,
where $\hat{\bf{\varphi}}$ is the unit vector in
azimuthal direction. Outside the wire ($r>a$) one has simply
$H_0(r) = I_{a0}/2\pi r$,
irrespective of the material properties and of $\omega$.
Inside this wire ($r \le a$)  one has
  \begin{eqnarray}  
  H_0(r) &=& {I_{a0} \over 2\pi a} \, {I_1(r/\lambda)
  \over I_1 (a/\lambda) } ~~~~=\,  \lambda^2 \,
  {\partial j_0(r) \over \partial r}  \,,
      \\
  j_0(r) &=& {I_{a0} \over \pi a^2}\,  {I_0(r/\lambda) \over
  I_1 (a/\lambda)}\, {a\over 2 \lambda} = {1 \over r}
  {\partial [r H_0(r)] \over \partial r}  \,,
  \end{eqnarray}
where $I_0(x)$ and $I_1(x)$ are Bessel functions. When $\lambda$
is real, Eqs.~(17) and (18) are well known London solutions,
and with imaginary $\lambda^2 = (\rho / i\omega \mu_0)$, or
complex $\lambda^{-1} = (1 + i)(\omega \mu_0 /2\rho)^{1/2} =
(1 + i)/\delta$, they describe the skin effect in Ohmic
wires with skin depth $\delta = (2\rho / \omega \mu_0)^{1/2}$.

  The complex resistance $Z= i\omega L +R$ and self-inductance
$L$ of this cylindrical wire may be obtained from the energy
balance \cite{1}.  In terms of real $I_a(t)$ and $E_a(t)$, the
input power $I_aE_a l$ is the sum of the temporal change of the
magnetic field energy outside the wire,
$(d/dt) \mu_0 l \int_a^{r_{\rm \!out}} 2\pi r (I_a/2\pi r)^2 dr =
 (d/dt) L_e I_a^2 /2$ defining the ``external inductance"
  \begin{eqnarray}  
   L_e= {\mu_0 l\over 2\pi} \ln {r_{\rm \!out} \over a} \approx
        {\mu_0 l\over 2\pi} \ln {l \over a}
  \end{eqnarray}
with outer cut-off radius $r_{\rm \!out} \approx l$, and the power
dissipated or stored inside the wire (in the magnetic field
and kinetic energy of the Cooper pairs).
Using the Pointing vector, this power may be expressed in
terms of the surface values $E_s=E(a,t)$ and $H_s =H(a,t)$,
thus $I_aE_a l = (d/dt) L_e I_a^2/2 +H_s E_s \cdot
 2\pi a l$. Inserting here $H_s = I_a/2\pi a$ and
dividing by $I_a$ we obtain
  \begin{eqnarray}  
   E_a(t)\,l = L_e\, \dot I_a(t) + E_s(t)\, l \,.
  \end{eqnarray}
Writing the linear Eq.~(20) in complex notation with
 $E_{a0}= Z I_{a0}$, $E_s = \rho j_0(a) \exp(i\omega t)$,
 $j_0(a)$ from Eq.~(18), and $R_{\rm wire} = \rho l/\pi a^2$,
we find
  \begin{eqnarray}  
  Z=i\omega\, {\mu_0 l\over 2\pi } \ln{l \over a} +R_{\rm wire} \,
  {a\over 2\lambda}\, {I_0(a/\lambda) \over I_1(a/\lambda)} \,.
  \end{eqnarray}
Formula (21) applies both to superconducting wires with real
London $\lambda$ (and
$R_{\rm wire} = i\omega \mu_0 \lambda^2 l / \pi a^2$)
and to Ohmic wires with $\lambda=\delta/(1+i)$
making the second term complex.
For superconducting wires with $\lambda \gg a$ this yields a
kinetic inductance $L_k = \mu_0 \lambda^2 l / \pi a^2$.
For Ohmic wires with large $\delta \gg a $, expansion of (21)
adds a constant $\beta=1/4$ to $\ln(l/a)$ (from the ``inner
inductance") and a factor
 $\gamma = 1+a^4 / 48 \delta^4 \approx 1$ to $R_{\rm wire}$.
For small skin depth $\delta \ll a$, this constant is
$\beta = \delta/a$ and the factor is
$\gamma = R/R_{\rm wire} = a/\delta \gg 1 $, i.e., the
resistance $R$ is strongly enhanced by the skin effect.

  In conclusion, the dependences on $\Lambda$, $\rho$,
$\omega$, and $l$, of the inductance $L$ and resistance $R$
of superconducting and Ohmic strips with applied ac current
are obtained, Eqs.~(12) to (16), and the current and magnetic
field profiles are depicted.
As compared to cylindrical wires, in strips the
enhancement of $R$ by the {\it skin effect is much weaker},
$R/R_{\rm strip} =\gamma \approx 1 +0.2 \ln(0.3 a d/\delta^2)$
for small skin depth $\delta \ll a$ and $d < \delta$.
Superconducting strips have a kinetic inductance
$L_k \approx \mu_0 l \Lambda/2a = \mu_0 \lambda^2 l/2ad$.
The geometric inductance of both Ohmic and superconducting
strips is similar to that of cylindrical wires;
it is dominated by the length-dependent factor $\ln(2l/a)$
that originates from the magnetic field energy outside the
conductor and does not depend on material parameters.
This factor $\ln(2l/a)$ results from Eq.~(7)
without having to introduce an outer cut-off here. The strip
length $l$ enters when the integral kernel
$K(x,x')$, Eq.~(3), is derived \cite{6} by integration of
the 3D Laplacian kernel $1/4\pi |{\bf r - r}'|$ over $z'$,
assuming $z$ independence of $J$. But, of course, the exact
value of the cut-off length in the logarithm depends also
on the return path of the current.

I thank J.~R.~Clem, G.~P.~Mikitik, and E.~Zeldov for helpful
discussions.
  This work was supported by the German Israeli Research
Grant (GIF) No G-705-50.14/01.

          {}

\end{document}